# ANOMALOUS THERMAL CONDUCTIVITY OF SINGLE CRYSTAL $Cu_2Te_2O_5Cl_2$

Dedicated to Prof. R. Blinc on the occasion of his 70[th] birthday

*A. Smontara[+], A. Bilušić[+], Z. Jagličić\*, J. Dolinšek[&] and H. Berger[$]*

[+]*Institute of Physics, P. O. Box 304, HR-10001, Zagreb, Croatia*
\**Institute for Mathematics, Physics and Mechanics , SI-1000 Ljubljana, Slovenia*
[&]*Jozef Stefan Institute, P.O. Box 3000, 1001 Ljubljana, Slovenia*
[$]*Institut de physique de la matire complexe, EPFL, CH-1015 Lausanne, Switzerland*

**Abstract**

A strong increase of the thermal conductivity is observed at the phase transition ($T_c$ =18.2 K) in $Cu_2Te_2O_5Cl_2$ single crystal. This behavior is compared with that of spin-Peierls system $NaV_2O_5$, where similar experimental observation has been found, and the conventional spin-Peierls system $CuGeO_3$, where a modest kink in the thermal conductivity curve has been observed. The strong increase of the thermal conductivity at $T_c$ in $Cu_2Te_2O_5Cl_2$ could be partially attributed to the opening of the energy gap in the magnetic excitation spectrum evident from the magnetic susceptibility measurements. However, the main reason for the anomaly of the thermal conductivity could be explained by a strong spin-lattice coupling in this system, while in $NaV_2O_5$ it is a consequence of a charge ordering.

## 1. Introduction

During the last several years a number of low-dimensional quantum-mechanical systems, such as the spin-Peierls $CuGeO_3$, the charge-ordered system $NaV_2O_5$, and the low-dimensional magnetic compound $SrCu_2(BO_3)_2$ [1] were discovered. When they are cooled down to the low temperatures, a spin gap openes which separations nonmagnetic ground state from the spectrum of spin excitations. The opening of the spin gap in these systems is manifested by the unusual low-temperature physical properties. Recently, the group of materials with the spin gap was expanded with the copper telluride $Cu_2Te_2O_5Cl_2$ system [2]. The new member possess unique tetrahedral clusters of $Cu^{2+}$ with $S = 1/2$, and belongs to the quantum magnets with antiferromagnetic interaction [1].

In comparison with the $CuGeO_3$, $NaV_2O_5$ and $SrCu_2(BO_3)_2$ systems, there exist a limited number of experimental data pertaining to the specific features of the antiferromagnetic (AF) transition in $Cu_2Te_2O_5Cl_2$ [2-4]. One of the experimental techniques not yet applied to the study of AF transition in $Cu_2Te_2O_5Cl_2$ is the measurement of thermal conductivity, which can provide useful information on the interaction of elementary excitations in this compound. Here, we report thermal conductivity measurements, which show an unexpectedly strong effect simillar to that in $NaV_2O_5$, in contrast with a rather weak anomaly observed in the $CuGeO_3$ system. A possible interpretation of this effect is considered.

## 2. Experimental

The single crystal used in our measurements was grown by the usual halogen vapor transport technique, using $TeCl_4$ and $Cl_2$ as transport agents. The semi-transparent light-green ($Cu_2Te_2O_5Cl_2$) sample was grown as a needle-like single crystal with the apparent chain morphology. Its dimensions were 1.8x1.6x10.5 mm$^3$, and mass of 54.9 mg.

In order to characterize the sample, the magnetization $M$(T) measurement was performed in a broad temperature interval from 2 K – 300 K in the low magnetic field of 100 Oe. The sample was field cooled from the room temperature down to the lowest temperature reached and then its magnetization $M(T)$ measured with increasing temperature in the same applied field of 100 Oe. The measurement was performed with a SQUID magnetometer (MPMS, Quantum Design) at Institute for Mathematics, Physics and Mechanics, Ljubljana.

The thermal conductivity was measured between 8 K and 300 K at the Institute of Physics, Zagreb, employing an absolute steady-state heat flow technique. The thermal flux through the samples was generated by a 1 kΩ $RuO_2$ chip-resistor, glued to one end of the sample, while the other end was attached to a copper heat sink (IMI 7031 was used as a glue). For monitoring the temperature gradient across the sample, 25 μm chromel-constantan differential thermocouple was used; its contacts were glued by IMI 7031 directly to the sample.

## 2. Results and discussion

The results of the magnetic susceptibility measurements $\chi(T) = M(T)/H$ of the $Cu_2Te_2O_5Cl_2$ single crystal are shown in Fig. 1 in the temperature range 2 K - 50 K, and in Fig. 2 in a broader temperature range 2 K - 300 K. The measurements were performed with the long sample axis parallel to the measuring field. The positions of the susceptibility maxima coincide with the values reported previously on the powder samples [2, 3] and is in a good agreement with the recent ac susceptibility measurements on the single crystals [5]. Our results show that there is a sharp kink at $T_c = 18.2$ K followed by an exponential decay ($\chi(T) \sim e^{-\Delta/T}$, with $\Delta = 43.1$ K) down to the value characterizing the low temperature susceptibility saturation $\chi = 5.24 \times 10^{-3}$ emu/mol*Oe. Verifying the congruence of our results with those published earlier [2], we have found that the suggested susceptibility model which takes into account isolated tetrahedrons (Fig. 2, thin line) of ref. 2 describes our data very well. In fact, there is a satisfactory fit to the results for the $Cu_2Te_2O_5Cl_2$ sample (Fig. 2), using the values of the interaction parameters $J_1/k_B = J_2/k_B = 41.2$ K which are simillar to those identified earlier $J_1/k_B = J_2/k_B = 38.5$ K [2]. In addition, our magnetic susceptibility data do not show a sizeable Curie-like upturn down to 2 K as previously reported [2]. This is an indication of the high quality of the sample used in our investigation.

The thermal conductivity measurements were performed along the long sample axis, and on the same sample for which the magnetization had been measured. The result of the thermal conductivity of the $Cu_2Te_2O_5Cl_2$ sample is shown in Fig. 1 and Fig. 3. The thermal conductivity of the $Cu_2Te_2O_5Cl_2$ sample smoothly increases with decreasing temperature, and then saturates at temperatures below 32 K with the appearance of a shallow dip at $T_C$. At still lower temperatures, the thermal conductivity starts to increase sharply, reaching a few times higher value at 8 K (the lowest temperature reached in our experimental set-up). A behavior similar to the one described above was seen in the charge-ordered system $NaV_2O_5$ [1] (see Fig. 3).

The enhancement of the thermal conductivity below $T_C$ could be attributed to the opening of an energy gap in the magnetic excitation spectrum. The opening of a spin gap switches off the spin-phonon scattering for phonons with frequencies less than the spin gap, so that more phonons contribute to the thermal conductivity below $T_c$. However, the thermal conductiviy results in the pure spin system $CuGeO_3$ [1] indicate that the spin-phonon scattering is not sufficient to explain the huge increase in the thermal conductivity. Similarly, the reason of the hudge increase in the thermal conductivity of $NaV_2O_5$ does not lie in the spin-gap opening, but in the charge ordering [1]. Thus, we believe that the increase of the thermal conductivity below $T_c$ in the $Cu_2Te_2O_5Cl_2$ system is a consequence of the unexpectedly large spin-lattice coupling, which drastically suppresses phonon conduction above $T_c$, and is efficiently swept away below $T_c$. Taking into account the explanations of the thermal conductivity in the $CuGeO_3$ and $NaV_2O_5$ systems, other reasons could not be excluded and more investigations are needed.

## 3. Conclusion

Our magnetic and thermal conductivity measurements reveal that the $Cu_2Te_2O_5Cl_2$ single crystal undergoes a phase transition at low temperatures, as well as the presence of a strong spin-lattice coupling. Further studies are underway to understand the microscopic background of this system.

**Acknowledgement**

This work was supported in part by the Croatian Ministry of Science and Technology (projects no. 0035013) and Croatian-Slovenian bilateral project. One of the author (A.S.) is grateful to Ms. M. Herak for her technical assistance. The sample preparation in Lausanne was supported by the NCCR research pool MaNEP of the Swiss NSF.

**References**

1. P.Lemmens, G.Guntherodt and C.Gros, Physics Report **376**, 1-103 (2003), *and references therein*.
2. M. Johnsson *et al.,* Chem. Mater. **12**, 2853 (2000).
3. P.Lemmens *et al.,* Phys. Rev. Lett. **87**, 227201 (2001).
4. C. Gros *et al.,* Phys.Rev.B **67**, 174405 (2003).
5. M. Prester, et. al., cond-mat/0307647.

**Figure captions**

Figure 1. Thermal conductivity $\kappa(T)$ and magnetic $\chi(T)$ susceptibility of $Cu_2Te_2O_5Cl_2$ single crystal, measured on the same specimen. The inset shows $\delta\chi(T)/\delta T$ in $10^{-4}$ emu/mol. The arrows denote the transition temperature.

Figure 2. Magnetic susceptibility of a quite large ($m$ = 54,9 mg) $Cu_2Te_2O_5Cl_2$ single crystal, measured in the field of $H$ = 100 Oe. There is a sharp kink at $T_c$ = 18.2 K followed by the exponential susceptibility decay (see inset) down to the value characterizing the low temperature susceptibility saturation. Thin solid line plots the isolated-tetrahedra model [2, 3] taken with $J_1/k_B = J_2/k_B = 41.2$ K.

Figure 3. Temperature dependences of thermal conductivity $\kappa(T)$ in $Cu_2Te_2O_5Cl_2$ single crystal, stoichiometric $NaV_2O_5$ [1] and $SrCu_2(BO_3)_2$ [1], for comparison.

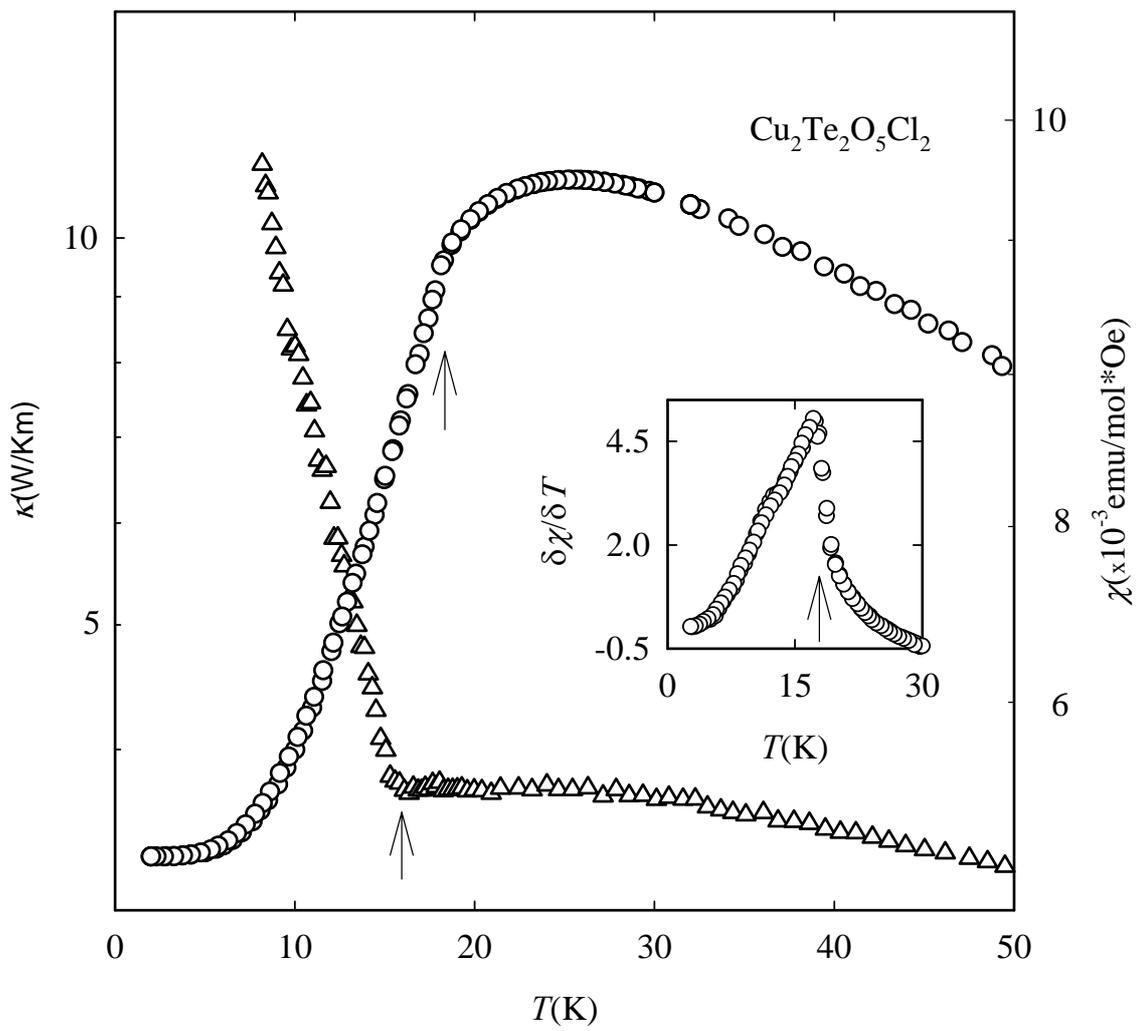

**Fig. 1**

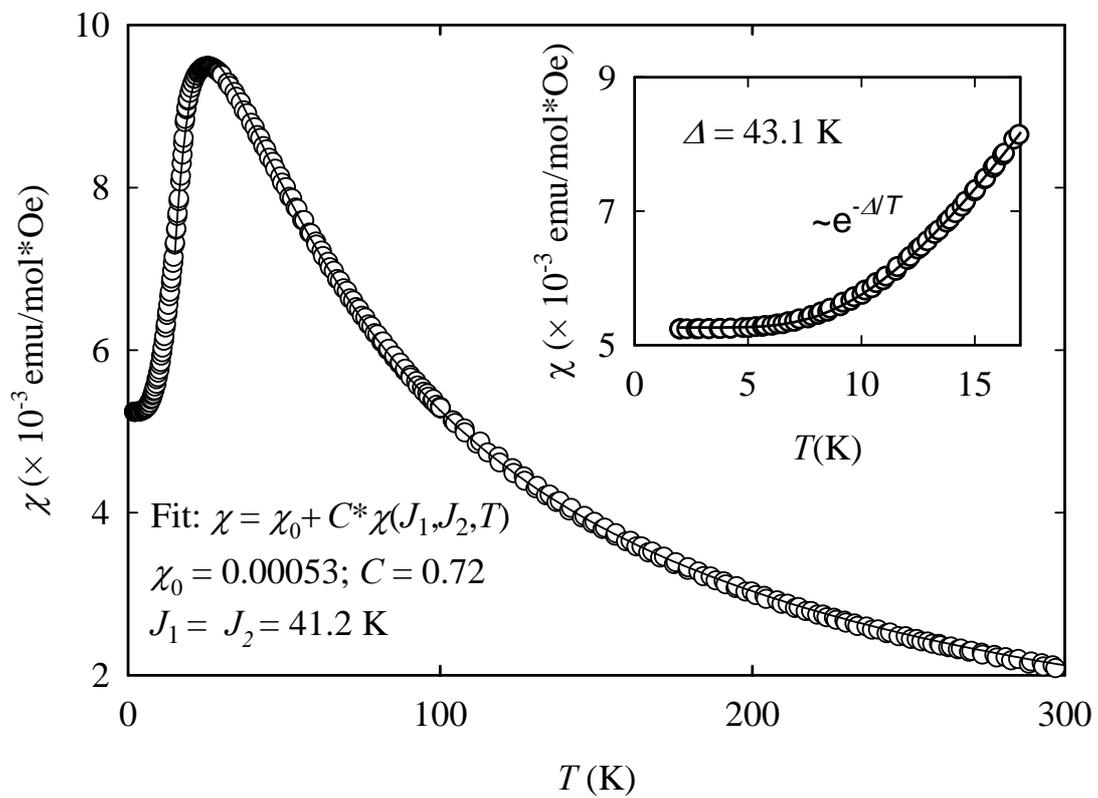

**Fig. 2**

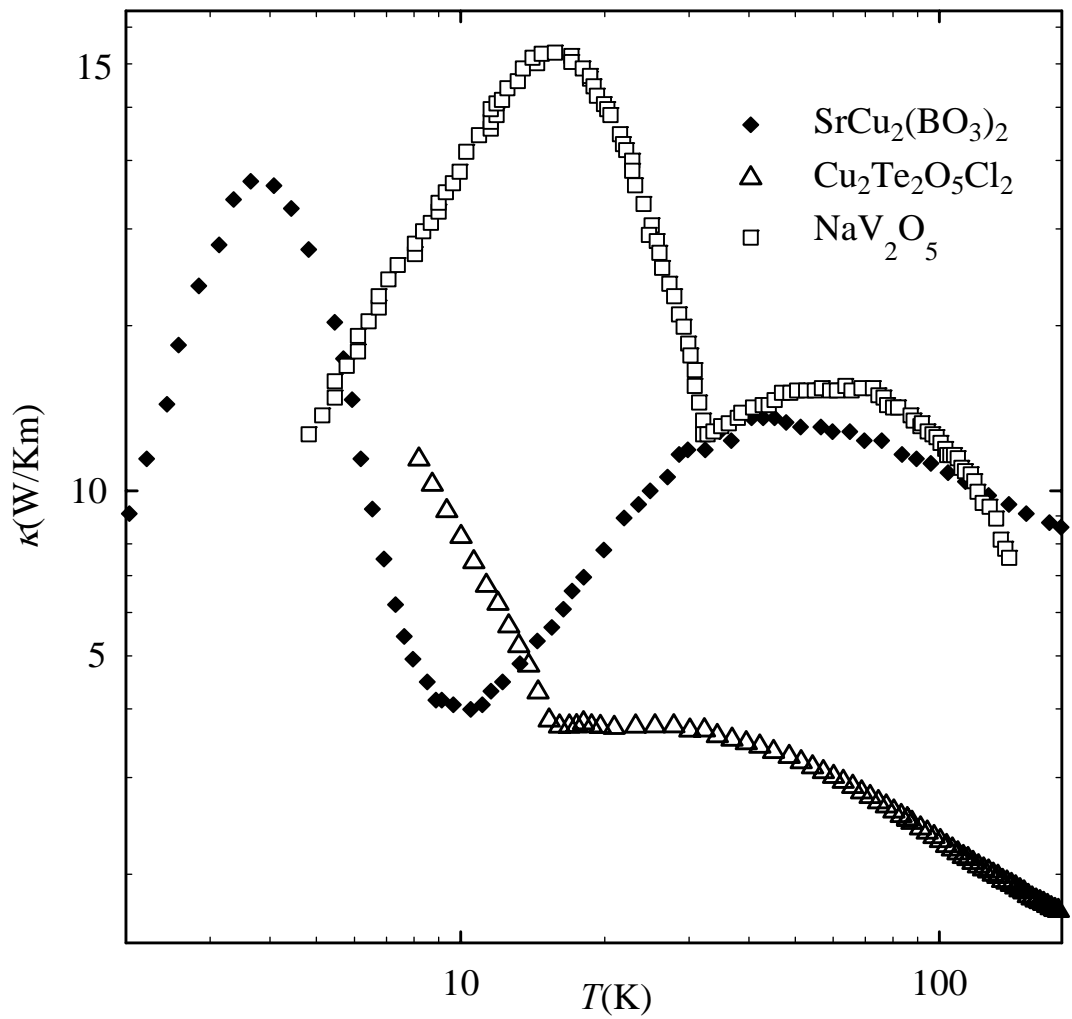

**Fig. 3**